%% file: main.tex
\documentclass[conference,10pt,letter]{IEEEtran}
\IEEEoverridecommandlockouts
\def\BibTeX{{\rm B\kern-.05em{\sc i\kern-.025em b}\kern-.08em
    T\kern-.1667em\lower.7ex\hbox{E}\kern-.125emX}}
\pdfoutput=1 %

\input{section/preamble/preamble}

\input{section/preamble/preamblelistingsconf}
\input{section/preamble/acronyms}
\input{section/preamble/macros}

\begin{document}

\title{
Benchmarking Large Language Models for Automated Verilog RTL Code Generation
}

\author{%
\IEEEauthorblockN{Shailja Thakur\IEEEauthorrefmark{1},  Baleegh Ahmad\IEEEauthorrefmark{1}, Zhenxing Fan\IEEEauthorrefmark{1}, Hammond Pearce\IEEEauthorrefmark{1}, \\Benjamin Tan\IEEEauthorrefmark{2}, Ramesh Karri\IEEEauthorrefmark{1}, Brendan Dolan-Gavitt\IEEEauthorrefmark{1},
Siddharth Garg\IEEEauthorrefmark{1}}
\IEEEauthorblockA{\IEEEauthorrefmark{1}New York University, \IEEEauthorrefmark{2}University of Calgary}
}

\maketitle

\begin{abstract}
Automating hardware design could obviate a significant amount of human error from the engineering process and lead to fewer errors. 
Verilog is a popular hardware description language to model and design digital systems, thus generating Verilog code is a critical first step. 
Emerging large language models (LLMs) are able to write high-quality code in other programming languages. 
In this paper, we characterize the ability of LLMs to generate useful Verilog. For this, we fine-tune pre-trained LLMs on Verilog datasets collected from GitHub and Verilog textbooks.
We construct an evaluation framework comprising test-benches for functional analysis and a flow to test the syntax of Verilog code generated in response to problems of varying difficulty. 
Our findings show that across our problem scenarios, the fine-tuning results in LLMs more capable of producing syntactically correct code (25.9\% overall).
Further, when analyzing functional correctness, a fine-tuned open-source CodeGen LLM can outperform the state-of-the-art commercial Codex LLM (6.5\% overall). %
Training/evaluation scripts and LLM checkpoints are available: \textcolor{linkcolor}{\url{https://github.com/shailja-thakur/VGen}}.

\end{abstract}


\input{section/01Introduction}

\input{section/02Background}

\input{section/03Datasets}

\input{section/04ModelTraining}
\input{section/05Evaluation}
\input{section/06Results}

\input{section/07Discussion}

\input{section/08Conclusions}

\bibliographystyle{IEEEtran}
\bibliography{lit/benhamram}

\end{document}

%% file: section/preamble/preamble.tex
\usepackage[utf8]{inputenc}
\usepackage[most]{tcolorbox}

\usepackage{booktabs}

\usepackage{graphicx}
\usepackage{multirow}
\usepackage[noadjust]{cite} 
\usepackage{setspace}
\usepackage{amsmath}
\usepackage{amssymb} %
\usepackage{amsfonts} %
\usepackage{calligra} %
\usepackage{mathtools}
\usepackage{amsthm} %
\usepackage[hidelinks,bookmarks=false]{hyperref}
\usepackage{dblfloatfix} %
\usepackage{array}
\usepackage{enumitem}

\usepackage{pgfplots}
\usepackage{pgfplotstable}
\usepgfplotslibrary{statistics}
\usepackage{cleveref}

\makeatletter
\let\MYcaption\@makecaption
\makeatother
\usepackage[font=footnotesize]{subcaption}
\makeatletter
\let\@makecaption\MYcaption
\makeatother

\usepackage{comment}
\usepackage{listings}

\usepackage{circledsteps}

\usetikzlibrary{patterns}

\usepackage{blindtext}

\usepackage{acronym}

\usepackage{xcolor, colortbl}

\usepackage{float}
\usepackage{textcomp}
\definecolor{linkcolor}{RGB}{219, 48, 122}

\usepackage{algorithm}
\usepackage{algorithmic}
\usepackage{url}

\usepackage[all=normal, floats=tight,indent=tight]{savetrees}
\usepackage{pifont}
\usepackage{framed}
\usepackage{soul}
\usepackage{multirow}
\usepackage{array}
\newcolumntype{L}[1]{>{\raggedright\let\newline\\\arraybackslash\hspace{0pt}}m{#1}}
\newcolumntype{C}[1]{>{\centering\let\newline\\\arraybackslash\hspace{0pt}}m{#1}}
\newcolumntype{R}[1]{>{\raggedleft\let\newline\\\arraybackslash\hspace{0pt}}m{#1}}
\newcolumntype{H}{>{\collectcell\lstinline}l<{\endcollectcell}}
\usepackage{url}

\newcommand*\circled[1]{\tikz[baseline=(char.base)]{
            \node[shape=circle,draw,inner sep=0.7pt] (char) {#1};}}

%% file: section/preamble/acronyms.tex
\acrodef{CPS}{Cyber-Physical System}
\acrodef{IoT}{Internet of Things}
\acrodef{HDL}{hardware description language}
\acrodef{CAD}{Computer-Aided Design}
\acrodef{EDA}{Electronic Design Automation}
\acrodef{HPC}{High-Performance Computing}
\acrodef{DL}{deep learning}
\acrodef{ML}{machine learning}
\acrodef{NLP}{natural language processing}
\acrodef{IC}{Integrated Circuit}
\acrodef{CWE}[CWE]{Common Weakness Enumeration}
\acrodef{CVE}[CVE]{Common Vulnerabilities and Exposures}
\acrodef{LLM}[LLM]{large language model}
\acrodef{NMT}[NMT]{neural machine translation}
\acrodef{NLP}[NLP]{natural language processing}

%% file: section/preamble/macros.tex
\newcommand{\ignore}[1]{{}}

\newcommand{\squishlist}{
	\begin{list}{$\bullet$}
		{ \setlength{\itemsep}{0pt}
			\setlength{\parsep}{1pt}
			\setlength{\topsep}{1pt}
			\setlength{\partopsep}{0pt}
			\setlength{\leftmargin}{0.9em}
			\setlength{\labelwidth}{1.5em}
			\setlength{\labelsep}{0.4em} } }
	\newcommand{\squishend}{
	\end{list}  }

\definecolor{graphFirst}{RGB}{2,136,209} %
\definecolor{graphSecond}{RGB}{211,47,47} %
\definecolor{graphThird}{RGB}{245,124,0} %
\definecolor{graphFourth}{RGB}{56,142,60} %
\definecolor{graphFifth}{RGB}{81,45,168} %
\definecolor{graphSixth}{RGB}{69,90,100} %
\definecolor{graphSeventh}{RGB}{251,192,45} %
\definecolor{backgroundSecond}{RGB}{239,154,154} %
\definecolor{backgroundThird}{RGB}{255,204,128} %
\definecolor{backgroundFourth}{RGB}{165,214,167} %
\definecolor{backgroundFifth}{RGB}{179,157,219} %
\definecolor{backgroundSixth}{RGB}{176,190,197} %
\definecolor{backgroundSeventh}{RGB}{255,245,157} %

%% file: section/01Introduction.tex
\section{Introduction}
\label{sec:intro}
State-of-the-art hardware design flows use \acp{HDL} such as Verilog and VHDL to specify hardware architectures and behaviors. %
However, the process of writing HDL code is time-consuming and bug-prone~\cite{dessouky_hardfails_2019}. 
As design complexity grows, there is a need to reduce design costs and developer effort during hardware specification. High-level synthesis tools enable developers to specify functionality in languages like C but come at the expense of hardware efficiency. A promising new approach is the use of \acp{LLM}~\cite{chen_evaluating_2021} 
to \emph{automatically} generate code from natural language specifications. \acp{LLM} are successful in  generating code in languages like C and Python. Their use in generating HDL code requires study. %

\acp{LLM} are deep neural networks, typically based on transformer architectures, that aim to model the underlying distribution of a natural or structured language corpus. Given a sequence of words (or ``tokens") \acp{LLM} predict a distribution over the next word/token. Used in a loop, \acp{LLM} can complete paragraphs in English starting  with the first sentence, or code from comments or initial lines of code. 

We undertake the first comprehensive evaluation of the syntactic and functional correctness of synthesizable Verilog code generated by both open-source and commercial \acp{LLM}. 
There are several challenges. 
First, baseline LLMs, including GitHub Copilot which ostensibly generates code in many programming languages including Verilog, frequently fail syntax, synthesis, and functional checks~\cite{pearce_asleep_2022}. 
Fine-tuning \acp{LLM} on a Verilog corpus can help, 
but requires a large dataset of Verilog code which is lacking. 
Prior work trained models on template-generated hardware specifications and corresponding Verilog, but this is time-consuming and does not generalize to unseen problems~\cite{pearce_dave_2020}. 
Finally, test problems and methods to evaluate the syntactic and functional correctness of LLM-generated code on a large scale are lacking.

\begin{figure}[!t]
    \centering
    \includegraphics[width=1\linewidth]{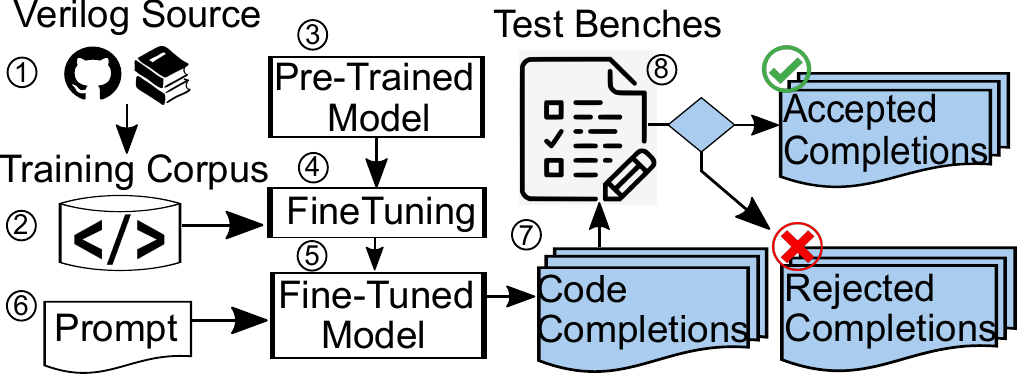}
    \caption{Experimental Evaluation of LLM Verilog Completions}
    \label{fig:system_overview}
\end{figure}

Our paper contributes the following. (1) By consolidating available open-source Verilog code as well as a broad search of textbooks about the Verilog \ac{HDL}, we create (to the best of our knowledge) the largest training corpus of Verilog code yet used for training \acp{LLM}. 
(2) Using this corpus, we examine fine-tuning five different pre-trained \acp{LLM} models with parameter counts ranging from 345M to 16B, producing five new fine-tuned models specialized for Verilog.
(3) To evaluate the efficacy of the models and determine the effect of the parameter sizes, we design a set of Verilog coding problems with varying difficulty levels, 
and corresponding test benches to test the functional correctness of generated code.

\autoref{fig:system_overview} illustrates our experimental platform for studying the impact of parameters such as temperature, number of sequences generated per problem, and number of LLM parameters. 
\autoref{sec:data} discusses creating the training data from GitHub and PDFs of Verilog textbooks \circled{1} with pre-processing \circled{2} and the five pre-trained LLMs \circled{3} that we fine-tune \circled{4} for completing Verilog code \circled{5}. 
\autoref{sec:evaluation} explains the evaluation setup, including our hand-designed prompts \circled{6}. 
\autoref{sec:Results} presents our results from generating code suggestions \circled{7} and evaluating with an analysis pipeline that compiles the Verilog and checks it against unit tests \circled{8}. 
\autoref{sec:discussion} discusses how our evaluation shows that the largest code-based \acp{LLM} (i.e., CodeGen-16B) fine-tuned on our Verilog corpus outperforms all other evaluated \acp{LLM}. Qualitatively, our best-performing \acp{LLM} can generate functioning code for challenging problems.

%% file: section/02Background.tex
\section{Background and Related Work} \label{sec:background}
\subsection{Background}
Transformer-based deep neural networks~\cite{vaswani_attention_2017} have demonstrated impressive ability in myriad domains, including language-related tasks.
Inputs to \acp{LLM} are in the form of tokens--- a set of common character sequences where each has a unique numeric identifier using a byte pair encoding~\cite{gage_new_1994}. 
Given a sequence of tokens as an input prompt, the \ac{LLM} outputs a probability distribution over the vocabulary for the next token given the prompt. A token is picked from this distribution, often the most likely token, appended to the prompt, and this sequence is fed back to the \ac{LLM}, yielding a new token. This is repeated to generate a \textit{completion}, a sequence of tokens that completes the input prompt.

\acp{LLM} for code are trained %
on a corpus of code in a target programming language or sometimes even on a 
mix of source code files in various languages. Dataset sizes can often be on the order of hundreds of gigabytes. Prompts for these \acp{LLM} can be in the form of comments, code snippets, or both. An \ac{LLM} trained on a mix of programming languages will often (implicitly) infer the language from the prompt.  

LLMs are  expensive to train from scratch due to their large datasets and massive parameter counts. However, pre-trained LLMs can be specialized for a user task by fine-tuning them on a specialized dataset. Fine-tuning is significantly faster than training from scratch because it only requires a small number of training epochs. Several LLMs pre-trained for both natural language and code either make the weights available, like NVIDIA's MegatronLM~\cite{shoeybi_megatron-lm_2020} or Salesforce's CodeGen models~\cite{nijkamp_conversational_2022}, or provide fine-tuning through an API, like AI21studio's Jurassic-1 (J1) models.\footnote{\url{{https://studio.ai21.com/docs/jurassic1-language-models/\#general-purpose-models}}}%

\subsection{Prior Work}
Programming is a challenging task, given the need for human designers to interpret and transform natural language specifications into programming structures.
This motivates the use of \ac{NLP} to transform language to code~\cite{mihalcea_nlp_2006}. Hardware design using Verilog HDL is similar to programming. Prior work explored \ac{NLP} techniques for %
generating assertions~\cite{harris_glast_2016}, albeit on a small scale. Pearce et al. trained \texttt{DAVE}, a small \ac{LLM} to produce Verilog snippets from template-based natural language descriptions for a limited set of functions~\cite{pearce_dave_2020}. 
GitHub's Copilot was evaluated for security bugs produced during out-of-the-box Verilog completions~\cite{pearce_asleep_2022} and was found to be lacking. 
This study is a large-scale exploration of the capabilities of \acp{LLM} across more design tasks using an automated evaluation framework.
There is no open dataset to train and evaluate LLMs on writing Verilog.

%% file: section/03Datasets.tex
\section{\ac{LLM} Training}
\label{sec:data}
In this section, we describe our method for training (or fine-tuning) \ac{LLM} models for Verilog code generation.
We begin by describing our curated Verilog datasets, followed by the \ac{LLM} architectures and the method for fine-tuning.

\subsection{Verilog Training Corpus}

Our primary Verilog training corpus comes from open-source Verilog code in public GitHub repositories. Additionally, we also created a dataset of text from Verilog textbooks to understand whether that further improved \ac{LLM} performance.

\paragraph{GitHub Corpus}  
We use Google BigQuery to gather Verilog repositories from GitHub, where it has a snapshot of over 2.8 million repositories. We use a query that looks for keywords such as ``Verilog" and files with  `.v' extension.
We de-duplicated files (using MinHash and Jaccard similarity metrics~\cite{yan_privmin_2017}) 
and filtered files by keeping `.v' files that contain at least one pair of \texttt{module} and \texttt{endmodule} statements. %
Finally, we filtered large files (number of characters $\geq$ 20K). %
The training corpus from GitHub yielded $\sim$50K files with a size of $\sim$300~MB. %

\paragraph{Verilog Books Corpus} We downloaded 70 Verilog-based textbooks from an online e-library in PDF format, then extracted text using the Python-based tool pymuPDF which uses optical character recognition to extract text. Depending on the quality of the PDF, the text quality varies. %
We cleaned the text by filtering irrelevant passages (e.g., index, preface, and acknowledgments) and used regular expressions to check high-level syntax of Verilog snippets from the surrounding prose, then use an overlapping sliding window on the filtered text corpus to produce training examples.
The final Verilog corpus of textbook-extracted and GitHub code had a size of 400~MB.

\subsection{Baseline LLM Architectures}
Table \ref{tbl:llm-architecture} shows the \acp{LLM} used in our study and summarizes design parameters including the number of layers, heads, embedding size (head dimension), context length, and the data source (natural language (NL) and/or code).
Since code-davinci-002 is derived from GPT-3~\cite{chen_evaluating_2021}, its architecture is the same as GPT-3. Its exact parameters are not known, so we leave these as \textit{NA} in the table.

\begin{table}[!t]
\caption{Baseline LLM architectures used in our study.}
\footnotesize
\resizebox{\linewidth}{!}{
    \begin{tabular}{L{3cm}C{0.8cm}C{0.8cm}C{0.8cm}C{1cm}}
    
        \toprule
        Model-Parameters / Pre-Training Data  & Layers & Heads & Embed. & Context Length\\
        \midrule
        MegatronLM-355M~\cite{shoeybi_megatron-lm_2020} / NL~\cite{devlin_bert_2019, radford_language_2019} & 24 & 16 & 64 & 1024 \\
        \midrule
        J1-Large-7B$^1$ %
        / NL~\cite{brown_language_2020} & 32 & 32 & 128 & 4096 \\
        \midrule
        CodeGen-2B~\cite{nijkamp_conversational_2022} / NL~\cite{gao_pile_2020}, Code
        & 32 & 32 & 80 & 2048 \\
        CodeGen-6B~ /  NL~\cite{gao_pile_2020}, Code & 33 & 16 & 256 & 2048\\
        CodeGen-16B / NL~\cite{gao_pile_2020}, Code & 34 & 24 & 256 & 2048\\
         \midrule
        code-davinci-002~\cite{chen_evaluating_2021} /  NL~\cite{brown_language_2020}, Code & NA & NA & NA &  8000\\
        \bottomrule
    \end{tabular}
    }
\vspace{0.5em}
\label{tbl:llm-architecture}
\end{table}

%% file: section/04ModelTraining.tex
\label{sec:training}

\subsection{LLM fine-tuning}
Five LLMs from~\autoref{tbl:llm-architecture} are fine-tuned on our Verilog training datasets. Training the CodeGen LLMs was challenging due to the large number of parameters. At 16-bit precision, CodeGen-16B's parameters alone occupy 30~GB of GPU memory; fine-tuning additionally requires sufficient GPU memory to store optimizer states and intermediate calculations, requiring around 250GB across multiple GPUs.
We use model and data parallelism and strategies for sharding the optimizer states across GPUs by basing our implementation on DeepSpeed.\footnote{\url{https://huggingface.co/docs/transformers/main_classes/deepspeed}}  %
We set the training hyperparameters to their defaults.
The CodeGen LLMs (2B, 6B, 16B) are fine-tuned for 1 epoch on an HPC cluster with two RTX8000s, four RTX8000s, and three A100s, and training completes in two, four, and six days, respectively. 
Megatron-LM is fine-tuned for 9 epochs using one RTX8000 for 15 hours using the default configuration~\cite{shoeybi_megatron-lm_2020}. We use the commercial off-the-shelf AI21 studio for fine-tuning J1-Large. 

%% file: section/05Evaluation.tex
\section{LLM Evaluation Setup}
\label{sec:evaluation}

The input to an LLM is a prompt from the problem set in \autoref{subsubsec:problem-set}. 
The \ac{LLM}-produced code completions on the problem are then truncated at keywords \texttt{end} and \texttt{endmodule}, and checked for compilation and functional correctness.
We use Icarus Verilog %
v11.0 for compiling the Verilog. %

\subsection{The Problem Set}
\label{subsubsec:problem-set}
The problem set has 17 Verilog problems inspired by classroom exercises that our team is familiar with and examples from the HDLBits website.\footnote{\url{https://hdlbits.01xz.net/wiki/Problem_sets}} %
We assign a difficulty level for each problem as shown in \autoref{tbl:problem_set}. The problem set covers designs spanning combinational and sequential logic. 
It has three finite state machines with different requirements, operations such as permutation, shift left, and rotate, and basic blocks like a MUX, RAM, LFSR, adders, and counters. 

\begin{table}[ht]
\caption{Problem set}
\footnotesize
\centering
\begin{tabular}{l l l} 
 \hline
 Prob. \# & Difficulty & Description \\ 
 \hline
 1 & Basic & A simple wire \\ 
 2 & Basic & A 2-input and gate \\
 3 & Basic &  A 3-bit priority encoder \\
 4 & Basic & A 2-input multiplexer \\
 5 & Intermediate & A half adder  \\
 6 & Intermediate & A 1-to-12 counter  \\
 7 & Intermediate & LFSR with taps at 3 and 5 \\
 8 & Intermediate & FSM with two states \\
 9 & Intermediate & Shift left and rotate \\
 10& Intermediate & Random Access Memory\\
 11& Intermediate & Permutation  \\
 12& Intermediate & Truth table \\
 13& Advanced & Signed 8-bit adder with overflow \\
 14& Advanced & Counter with enable signal \\
 15& Advanced & FSM to recognize `101' \\
 16& Advanced & 64-bit arithmetic shift register \\
 17& Advanced & ABRO FSM$^{*}$ %
 \\
 \hline
\end{tabular}
\\ $^{*}$\textit{from Potop-Butucaru, Edwards, and Berry's ``Compiling Esterel''}

\vspace{1em}

\label{tbl:problem_set}
\end{table}

Examples for basic, intermediate and advanced problems are shown in Fig.~\ref{fig:basic-example}-\ref{fig:advanced-example} respectively. They are obtained using CodeGen-16B-FT and edited for visualization and brevity.

\begin{figure}
\centering
\begin{subfigure}[b]{0.95\linewidth}
\begin{lstlisting}[language=verilog]
// This is a 3-bit priority encoder. It outputs the position of the first high bit.
module priority_encoder(input[2:0] in,output reg[1:0] pos); 
// If none of the input bits are high (i.e., input is zero), output zero.
// assign the position of the highest bit of in to pos.
\end{lstlisting}
\vspace{-3mm}
\caption{Prompt}
\label{fig:basic-example-prompt}
\end{subfigure}

\begin{subfigure}[b]{0.95\linewidth}
\begin{lstlisting}[language=verilog]
always @(in) 
if (in == 0)  pos <= 2'h0;
else if (in[0])  pos <= 2'h0;
else if (in[1]) pos <= 2'h1;
else pos <= 2'h2;
endmodule
\end{lstlisting}
\vspace{-3mm}
\caption{Correct completion}
\label{fig:basic-example-correct_completion}
\end{subfigure}

\begin{subfigure}[b]{0.95\linewidth}
\begin{lstlisting}[language=verilog, linebackgroundcolor={\ifnum\value{lstnumber}>3
                \ifnum\value{lstnumber}<7
                    \color{pink}
                \fi
            \fi }]
always @(in) begin 
case(in)
 3'b000: pos=2'b00;
 3'b001: pos=2'b01;
 3'b010: pos=2'b10;
 3'b011: pos=2'b11;
default: pos=2'b00;
. . . 
\end{lstlisting}
\vspace{-3mm}
\caption{Incorrect completion. Positions are offset by 1.}
\label{fig:basic-example-incorrect_completion}
\end{subfigure}

\vspace{-3.5mm}
\caption{Basic example - Problem 3: A 3-bit priority encoder}
\label{fig:basic-example}
\end{figure}

\begin{figure}
\centering
\begin{subfigure}[b]{0.95\linewidth}
\begin{lstlisting}[language=verilog]
// This is a counter that counts from 1 to 12
module counter(input clk,input reset,output reg[3:0] q); 
\end{lstlisting}
\vspace{-3mm}
\caption{Prompt}
\label{fig:intermediate-example-prompt}
\end{subfigure}

\begin{subfigure}[b]{0.95\linewidth}
\begin{lstlisting}[language=verilog]
always @(posedge clk) begin
if(reset) q <= 4'd1;
else begin 
if(q == 4'd12) q <= 4'd1;
else q <= q + 4'd1;
\end{lstlisting}
\vspace{-3mm}
\caption{Correct completion}
\label{fig:intermediate-example-correct_completion}
\end{subfigure}

\begin{subfigure}[b]{0.95\linewidth}
\begin{lstlisting}[language=verilog, linebackgroundcolor={\ifnum\value{lstnumber}>3
                \ifnum\value{lstnumber}<6
                    \color{pink}
                \fi
            \fi }]
always @ (posedge clk) begin
if(reset) q <= 4'd1;
else begin
q <= q + 4'd1;
end end endmodule
\end{lstlisting}
\vspace{-3mm}
\caption{Incorrect completion. Counter does not stop at 12.}
\label{fig:intermediate-example-incorrect_completion}
\end{subfigure}

\vspace{-3.5mm}
\caption{Intermediate example - Problem 6: A 1 to 12 counter}
\label{fig:intermediate-example}
\end{figure}

\begin{figure}[t]
\centering
\begin{subfigure}[b]{0.95\linewidth}
\begin{lstlisting}[language=verilog]
// This is an FSM 
// It outputs 1 when 1 is received for signals a and b irrespetive of their order, either simultaneously or non-simultaneously.
module abro(input clk,input reset,input a,input b,output z);
parameter IDLE = 0, SA = 1, SB = 2, SAB = 3;
reg [1:0] cur_state,next_state;
//Update state or reset on every clock edge
//Output z depends only on the state SAB
//The output z is high when cur_state is SAB
//cur_state is reset to IDLE when reset is high. Otherwise, it takes value of next_state.
//Next state generation logic:
//If cur_state is IDLE and a and b are both high, state changes to SAB
//If cur_state is IDLE,and a is high,state changes to SA
//If cur_state is IDLE,and b is high,state changes to SB
//If cur_state is SA,and b is high,state changes to SAB
//If cur_state is SB,and a is high,state changes to SAB
//If cur_state is SAB, state changes to IDLE
\end{lstlisting}
\vspace{-3mm}
\caption{Prompt}
\label{fig:advanced-example-prompt}
\end{subfigure}

\begin{subfigure}[b]{0.95\linewidth}
\begin{lstlisting}[language=verilog]
// Implements an FSM in Verilog
always @(posedge clk or posedge reset) begin
if(reset) cur_state <= IDLE;
else cur_state <= next_state; end
always @(cur_state or a or b) begin
case(cur_state)
IDLE: begin
if(a && b) next_state = SAB;
else if (a) next_state = SA;
else if (b) next_state = SB; end
SA: begin
if(b) next_state = SAB;
else next_state = SA; end
SB: begin
if(a) next_state = SAB;
else next_state = SB; end
SAB: begin
next_state = IDLE; end
default:
next_state = IDLE; endcase end
// Output logic:
// Output z is high when cur_state is SAB
// Output z is high when cur_state is IDLE and a and b are both high
// Output z is low otherwise.
assign z = (cur_state == SAB || (cur_state == IDLE && a && b));
endmodule
\end{lstlisting}
\vspace{-3mm}
\caption{Correct completion}
\label{fig:advanced-example-correct_completion}
\end{subfigure}

\begin{subfigure}[b]{0.95\linewidth}
\begin{lstlisting}[language=verilog, linebackgroundcolor={\ifnum\value{lstnumber}>1
                \ifnum\value{lstnumber}<4
                    \color{pink}
                \fi
            \fi }]
... SAME AS (b) TILL LINE 24 ...
assign z = (cur_state == IDLE && a && b) || (cur_state
 == IDLE && a);
endmodule
\end{lstlisting}
\vspace{-3mm}
\caption{Incorrect completion. Output is not assigned to state SAB.}
\label{fig:advanced-example-incorrect_completion}
\end{subfigure}

\vspace{-3.5mm}
\caption{Advanced example - Problem 17}
\label{fig:advanced-example}
\end{figure}

   \begin{figure}[h]
    \centering
    \begin{subfigure}[t]{0.97\linewidth}
    \begin{lstlisting}[language=verilog,linebackgroundcolor={\ifnum\value{lstnumber}>4
                \ifnum\value{lstnumber}<9
                    \color{yellow}
                \fi
            \fi 
            \ifnum\value{lstnumber}>8
                \ifnum\value{lstnumber}<16
                    \color{lightgray}
                \fi
            \fi}]
// This is a finite state machine that recognizes the sequence 101 on the input signal x. 
module adv_fsm(input clk, input reset, input x, output z); 
reg [1:0] present_state, next_state;
parameter IDLE=0, S1=1, S10=2, S101=3;
// output signal z is asserted to 1 when present_state is 
// S101
// present_state is reset to IDLE when reset is high, 
// otherwise it is assigned next state
// if present_state is IDLE, next_state is assigned S1 if 
// x is 1, otherwise next_state stays at IDLE
// if present_state is S1, next_state is assigned S10 if
// x is 0, otherwise next_state stays at IDLE 
// if present_state is S10, next_state is assigned S101 if 
// x is 1, otherwise next_state stays at IDLE 
// if present_state is S101, next_state is assigned IDLE  
    \end{lstlisting}
    \end{subfigure}
    \vspace{-3mm}
    \caption{Varying the prompt details: Low, Medium and High. Problem 15.}
    \label{fig:prompt-details}
    \end{figure}

\subsection{Input parameters}

Each \ac{LLM} query has a prompt, a sampling temperature ($t$), and a number of completions/prompt ($n$). 

\noindent \textbf{Prompts:} %
    Three prompts with increasing detail i.e., low (L), medium (M), and high (H) are provided.
    Prompt L has an initial comment describing the function of the module and the module header with name and inputs/outputs with types. 
    Internal signals are also declared. 
    M includes L plus comments that describe the function using signal names. 
    H replaces and/or appends comments in M with more detail and resembles pseudo-code as opposed to a predominantly natural language specification. \autoref{fig:prompt-details} is an example for Problem 15. 
    L has no lines highlighted (the prompt is lines 1--4). 
    M includes L and lines highlighted yellow (the prompt is lines 1--8). 
    H includes M and lines in gray (the prompt is lines 1--15).

\noindent \textbf{Sampling temperature $(t)$:} A higher value means that the LLM takes more risks and yields more creative completions. We use $t \in \{0.1,0.3.0.5,0.7,1\}$.
    
\noindent \textbf{Completions per prompt $(n)$:} For each prompt, LLM generates $n$ completions where $n \in \{1,10,25\}$. For J1-Large, we skip $n=25$ because they do not support this value.

\noindent \textbf{max\_tokens:} The maximum number of tokens generated for each completion was set to 300 for all LLMs except J1-Large. For J1-Large the limit is 256. Nucleus sampling probability mass (\texttt{top\_p}) was set to the default value of 1.

\subsection{Test benches}
For each problem, we developed a test bench to check for functional correctness. 
The test benches exercise the designs for corner cases and are exhaustive for basic and some intermediate cases. 
For the remaining cases,
the test benches are analogous to unit tests. 
This keeps the time to evaluation reasonable. For example, for the RAM module, the data width is 8 and the address width is 6 in the prompt.  An exhaustive test bench requires $2^{14}$ test inputs and would take longer to simulate. 
In some cases, specifications in the prompts are ambiguous and thus can yield several correct responses. For example, 
when one does not specify whether a reset should be synchronous or asynchronous. %

%% file: section/06Results.tex
\section{LLM Evaluation and Results}
\label{sec:Results}
\subsection{Research Questions}
This study examines four research questions (RQs) from the point of view of quality of Verilog generation given the scenarios and test-benches from Section~\ref{sec:evaluation}:
\textbf{RQ1}. How well do `base' LLMs perform on the Verilog generation set?
\textbf{RQ2}. Does fine-tuning LLMs improve that performance?
\textbf{RQ3}. Are larger LLMs with more parameters better?
\textbf{RQ4}. Does variability in problem description impact quality and the number of correct completions?

\subsection{Results}
We measure the quality of the code generated by LLMs using problem sets described in~\autoref{sec:evaluation}. 
A scenario is a combination of problems across difficulties and description levels. 
We query the models with all prompt $\times$ $t$ $\times$ $n$ combinations. 
For fairness, we present each model's ``\textit{best results}'' by focusing on the completions generated with the $t$ for each model for which their completions were most successful at compiling and passing the functional tests (for each problem difficulty and description level). 
We present these \textit{best results} for $n=10$ in \autoref{tbl:compiled} and \autoref{tbl:results}. 
\autoref{tbl:compiled} shows the proportion of completions that compile and \autoref{tbl:results} shows the proportion of completions that pass functional tests, for the completions produced by a given temperature setting that resulted in the most successful completions for each scenario.
As in prior work~\cite{nijkamp_conversational_2022}, we characterize the model performance with the Pass@$k$ metric, where $k$ is the number of problems in a \textit{scenario} times $n$, the number of suggestions per problem. A higher Pass@$k$ indicates a relatively `better' result. 
For compilation (\autoref{tbl:compiled}), the Pass@$k$ metric reflects the proportion of completions that compile.
For functional tests, this metric is the fraction of the $k$ code samples that pass.

For interest, Table~\ref{tbl:results} reports the inference time for each \acp{LLM} query, including communication time with a remote server if required. %
Note that the results are after fine-tuning the model using the training corpus from GitHub only. We discuss the case for fine-tuning on GitHub and PDFs combined as an ablation study in the discussion. Fine-tuned CodeGen-16B LLM outperforms all LLMs. %
All fine-tuned LLMs outperform their pre-trained counterparts. %
[\textbf{Ans. RQ1 and RQ2}].   

\subsubsection{Completions vs. Temperature ($t$)}%
\begin{figure}[t]
    \centering
    \includegraphics[width=1.0\linewidth]{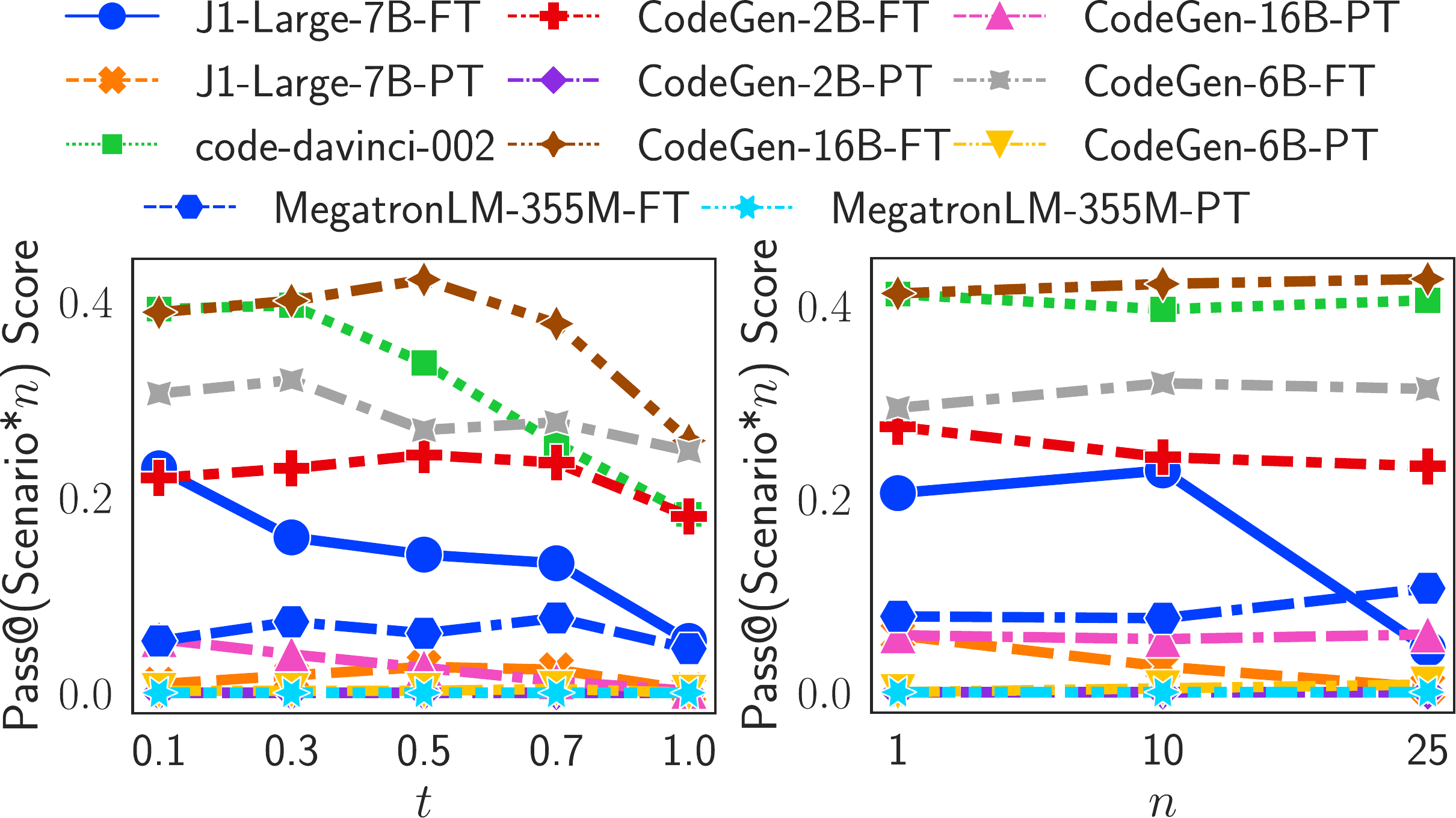}
    \vspace{-1.5em}
    \caption{Pass@(scenario*$n$) for scenarios passing test-benches across temperature ($t$) and completions per prompt ($n$). Higher is better.}
    \label{fig:t-n-change}
\end{figure}

\autoref{fig:t-n-change} summarizes the Pass@(\textit{scenario}*$n$) metric for our experiments sweeping temperature. %
Pass@(\textit{scenario}*10) has the highest value for $t=0.1$ and degrades exponentially with temperature. The LLM  generates accurate solutions at low temperatures
and accurate synthesizable codes are expected from a smaller number of candidates. 

\subsubsection{Completions vs. \# Completions/Prompt ($n$)}%
We study synthesis quality as a function of completions/prompt. 
The right-hand panel in \autoref{fig:t-n-change} shows the Pass@(\textit{scenario}*$n$) for all LLMs. Pass@(\textit{scenario}*$1$) is better than Pass@(\textit{scenario}*$10$). This improves as the  number of completions increases. %
This is the case because the number of candidate solutions at low temperatures increases, increasing the completions passing the test benches. $n=10$ is good for all problem difficulty levels.

\subsubsection{Completions vs. LLM Size} %
\autoref{fig:t-n-change} and \ref{fig:description-difficulty} show that LLMs with more parameters (CodeGen-16B, code-davinci-002) outperform LLMs with less parameters such as Megatron-355M and CodeGen-2B. These LLMs yield  more completions that pass test benches and more correct completions. [\textbf{Ans. RQ3}].
\subsubsection{Completions vs. Prompts} %
\begin{figure}[t]
    \centering
    \includegraphics[width=1.0\linewidth]{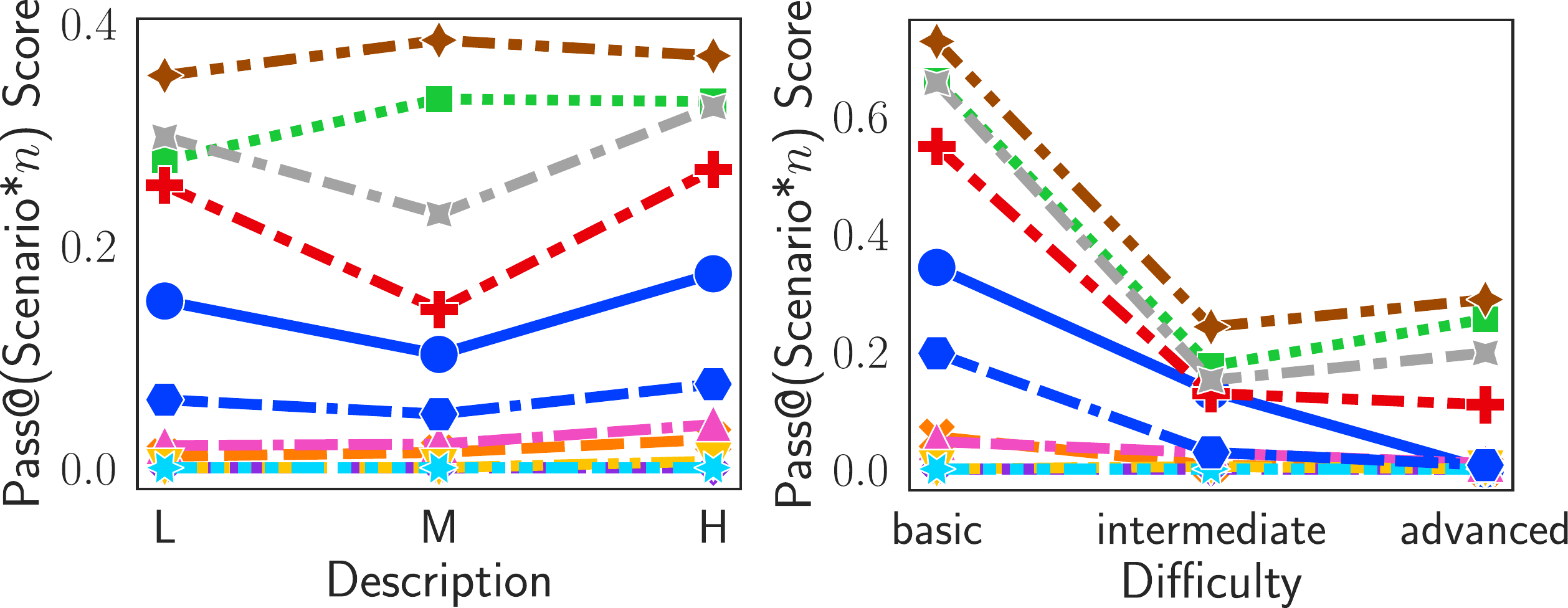}
    \vspace{-1.5em}
    \caption{Pass@(scenario*$n$) for scenarios passing test-benches across problem difficulties and description levels. Higher is better.}
    \vspace{-0.3em}
    \label{fig:description-difficulty}
\end{figure}

Prompt quality impacts the LLM generation quality. We study the effect of variations in the prompt description at two levels: How do difficulty of the prompt and the description of the prompt impact code completions. We use Pass@(\textit{scenario}*$10$) as the metric.
The right-hand side panel in \autoref{fig:description-difficulty} shows that the Pass@(\textit{scenario}*$10$) decreases with increasing prompt difficulty.
Simple problems such as AND are easy to translate to Verilog, as opposed to advanced problems such as LFSR. %
The left-hand side panel in \autoref{fig:description-difficulty} shows that the number of correct solutions decreases with terse prompts. [\textbf{Ans. RQ4}].

\begin{table}[!t]
\footnotesize
\centering
\caption{Pass@(scenario*$n$) at $n$=10 for compiled completions (Pass=Compiling), PT = Pre-trained, FT = Fine-Tuned. Bold reflects the (best) highest performance for that difficulty.}
\begin{tabular}{lcccc}
\toprule
\textbf{Model} & \textbf{Model Type} & Basic &  Intermediate &  Advanced \\
\midrule
\multirow{2}{*}{MegatronLM-345M} & PT & 0.000 &         0.000 &     0.000 \\
 & FT &  0.730 & 0.391 &     0.165 \\
 \midrule
\multirow{2}{*}{CodeGen-2B}  & PT    &  0.080 &         0.065 &     0.176 \\
   & FT   &  0.902 & 0.612 &     0.592 \\
\midrule
\multirow{2}{*}{CodeGen-6B} & PT      &  0.052 &         0.152 &     0.187 \\
& FT      &  \textbf{0.987} &   0.689 &     \textbf{0.599} \\
\midrule
\multirow{2}{*}{J1-Large-7B} & PT      &  0.182 &         0.176 &     0.108 \\
& FT      &  0.882 &    0.635 &     0.588 \\
\midrule
\multirow{2}{*}{CodeGen-16B} & PT     &  0.132 &         0.203 &     0.240 \\
& FT     &  0.942 &    \textbf{0.728} &     0.596 \\
\midrule
code-davinci-002 & PT   &  0.847 &         0.452 &     0.569 \\
\bottomrule
\end{tabular}
\label{tbl:compiled}
\end{table}

\begin{table*}
\centering
\caption{Pass@(scenario*$n$) at $n$ =10 for test bench passing completions (Pass=Passed Functional Tests), PT = Pre-trained, FT = Fine-Tuned. Bolded value in each test column reflects the (best) highest performance for that problem set and difficulty.}
\footnotesize
\begin{tabular}{lm{0.9cm}c|ccc|ccc|ccc}
\toprule
\multirow{3}{*}{\textbf{Model}} & \multirow{3}{1em}{\textbf{Model Type}}  & \multirow{3}{5em}{\textbf{Inference Time (s)}} & \multicolumn{3}{c}{\textbf{Basic}} & \multicolumn{3}{c}{\textbf{Intermediate}} & \multicolumn{3}{c}{\textbf{Advanced}}  \\
\cmidrule(lr){4-12}
        &  & &L & M & H & L & M & H & L & M & H  \\
\midrule
\multirow{2}{*}{MegatronLM-355M} & PT &   3.628    &   0.000 &          0.000 &          0.000 &                 0.000 &                 0.000 &                 0.000 &             0.000 &             0.000 &             0.000\\
 & FT &          0.175 &          0.170 &   0.591   &    0.245 &                 0.043 &                 0.018 &                 0.025 &             0.000 &             0.000 &             0.000 \\
 \midrule
 \multirow{2}{*}{CodeGen-2B} & PT &      1.478  &          0.000 &          0.000 &          0.000 &                 0.000 &                 0.000 &                 0.000 &             0.000 &             0.016 &             0.020\\
 & FT       &          0.665 & 0.835       &  0.350 &          0.630 &                 0.130 &                 0.092 &                 0.163 &             0.132 &             0.048 &             0.068 \\
 \midrule
 \multirow{2}{*}{CodeGen-6B} & PT  &  2.332     &          0.000 &          0.000 &          0.000 &                 0.000 &                 0.000 &                 0.013 &             0.000 &             0.000 &             0.000\\
  & FT       &          0.710 &    \textbf{1.000}    &  0.500 &          0.760 &                 0.135 &                 0.150 &                 0.168 &             \textbf{0.284} &             0.164 &             0.164\\
 \midrule

\multirow{2}{*}{J1-Large-7B} & PT   & 7.146      &          0.044 &          0.058 &          0.067 &                 0.000 &                 0.000 &                 0.021 &             0.000 &             0.000 &             0.000 \\
 & FT         &   2.029    &   0.388 &          0.283 &          0.342 &                 0.125 &                 0.075 &                 0.200 &             0.000 &             0.000 &             0.000 \\
\midrule

 \multirow{2}{*}{CodeGen-16B} & PT      &    2.835    &  0.000 &          0.085 &          0.055 &                 0.035 &                 0.003 &                 0.045 &             0.012 &             0.000 &             0.016\\
 & FT      &    1.994    &  0.745 &          \textbf{0.720} &          0.745 &                 \textbf{0.213} &                 \textbf{0.270} &                 \textbf{0.255} &             0.246 &             \textbf{0.290} &             0.294\\
 \midrule
code-davinci-002 & PT   &   3.885  &     0.520 &          0.685 &          \textbf{0.775} &                 0.175 &                 0.200 &                 0.150 &             0.156 &             0.184 &             \textbf{0.344}\\
\bottomrule
\end{tabular}
\vspace{-0.5em}
\label{tbl:results}
\end{table*}

%% file: section/07Discussion.tex
\section{Discussion and Limitations}
\label{sec:discussion}

As shown in \autoref{tbl:results}, fine-tuned LLMs generate code that compiles better when compared to the pre-trained LLMs. Using the best Pass@(\textit{scenario}*$10$) values, only $11.9\%$ of the completions generated by pre-trained LLMs compiled while $64.6\%$ of those by fine-tuned LLMs compiled. 
Thus, a designer may use these \acp{LLM} with text and/or pseudo-code to generate a syntactically-correct `skeleton' of a design, before then tweaking it to meet functional requirements.

We assess \acp{LLM}' code completion using the associated Verilog test-benches. 
These test-benches are comprehensive for the Basic problems, but as the problems become more complex, the test-benches cover only those behaviors fully specified in the problem comments.
As LLMs tend to provide similar responses when several completions per prompt are requested, the exact test-bench implementation can have a large impact on how many cases pass. 
We observe an example of this in the \acp{LLM}' responses to FSM problems 8, 15, and 17. Since the problem comments do not specify whether the reset is synchronous/asynchronous, the LLMs are free to produce any variation.
As such, for all problems, we verify whether an active-high reset results in the appropriate value at the output, but we do not test the asynchronous/synchronous corner case nor other similar edge conditions. %

Even the best performing LLM (CodeGen-16B (FT)) performed poorly for some problem sets. %
For any given problem, CodeGen-16B (FT) produced 540 completions, but for Problems 7 (LFSR) and 12 (Truth table), none of the completions passed, and for Problem 9 (Shift and Rotate), only one passed. We manually investigated the completions and observed that for Prob. \# 7, the LLMs had trouble concatenating the most significant bits with the feedback value. %
This was the problem in most cases and a better prompt might yield a correct result. This indicates the importance of creating the best prompt, pointing to prompt engineering as future work.
For Prob. \#9, completions  either do not cover all values of the shift or assign incorrect bit positions. For Prob. \#12, completions are close to the actual solution by using all input values in \texttt{assign} statements but fail to form correct expressions between input bits.
This suggests insufficient diversity in the training corpus.

Next, we study the impact of the training corpus 
on LLM fine-tuning. We conduct an ablation study using (a) CodeGen-16B fine-tuned with GitHub verilog repositories only and (b) CodeGen-16B fine-tuned with Verilog from Github and textbooks. The Pass@(\textit{scenario}*$10$) for (a) and (b) show that
option (b) is marginally better (1.4\%) than (a). This is the case because the Verilog corpus from PDFs adds more examples and this helps the LLM to generalize to Verilog.

%% file: section/08Conclusions.tex
\section{Conclusions}
This paper describes a new paradigm for automatically generating and verifying Verilog from \acp{LLM}. 
Using the presented Pass@(\textit{scenario}*$n$) values from Tables~\ref{tbl:compiled}-\ref{tbl:results}, pre-tuned LLMs produced completions that are functionally correct only $1.09\%$ of the time. This number increases to $27.0\%$ after tuning, showing a clear benefit to fine-tuning \acp{LLM} over a specific language.
The fine-tuned CodeGen-16B LLM was the most successful in completions with respect to functional correctness.
Overall it produced functionally correct code $41.9\%$ of time, whereas the commercially available state-of-the-art (non-fine-tuned) code-davinci-002 LLM 
produced functionally correct code $35.4\%$ of time. %

\label{sec:conclusions}

\section*{Acknowledgements}

This research work was supported in part by NSF Award 1553419, NSF Award 1646671, NSF Award 2039607, and ARO Award 77191NC. The opinions, findings, and conclusions, or recommendations expressed are those of the author(s) and do not necessarily reflect the views of any sponsors.